\documentclass[12pt]{article}
\usepackage{graphicx}

\begin{document}

\title{Dynamics of strangeness and collective flows in heavy-ion collisions near threshold energies}

\author{Zhao-Qing Feng \footnote{Corresponding author. Tel. +86 931 4969152. \newline \emph{E-mail address:} fengzhq@impcas.ac.cn (Z.-Q. Feng)}}
\date{}
\maketitle

\begin{center}
\small \emph{Institute of Modern Physics, Chinese Academy of Sciences, Lanzhou 730000, People's Republic of China}
\end{center}

\textbf{Abstract}
\par
Strangeness (K$^{0,+}$, $\Lambda$ and $\Sigma^{-,0,+}$) production in heavy-ion collisions near threshold energies has been investigated within the Lanzhou quantum molecular dynamics (LQMD) transport model. The kaon (anti-kaon)-nucleon and hyperon-nucleon potentials in dense nuclear matter are implemented in the model, which are computed from the chiral perturbation approach and the relativistic mean-field model, respectively. It is found that the in-medium potentials change the structure of transverse flow, and also affect the rapidity distributions and the inclusive spectra for strangeness production. The local temperature of the fire ball extracted from the kaon spectra of inclusive invariant cross sections is influenced by the kaon-nucleon potential. The stiffness of nuclear symmetry energy and the kaon-nucleon potential by distinguishing isospin effect play a significant role on the ratio of K$^{0}$/K$^{+}$, in particular at the subthreshold energies. The ratio of $\Sigma^{-}/\Sigma^{+}$ depends on the high-density symmetry energy, in which the $\Sigma$-nucleon potential has a neglectable contribution on the isospin ratio.

\bigskip

\section{Introduction}

Strange particles produced in relativistic heavy-ion collisions has attracted much attention over past several decades. The interaction of strange particles and baryons in dense nuclear matter is not well understood up to now, which can not be obtained from the data of exotic atoms \cite{Fr07}. Heavy-ion collisions in terrestrial laboratory provide a unique tool to study the in-medium properties of strangeness in dense nuclear matter and to extract the nuclear equation of state (EoS). It has obtained much progress for constraining the high-density information of isospin symmetric EoS both experimentally \cite{Fo07,Me07} and theoretically \cite{Ai85,Li97,Fu01,Ha06,Fe10} from kaon production in heavy-ion collisions. Kaons ($K^{0}$ and $K^{+}$) as a probe of EoS are produced in the high-density domain almost without subsequent reabsorption in nuclear medium. The influence of the kaon-nucleon potential on EoS is very small \cite{Fe11a}. The available experimental data from $K^{+}$ production already favored a soft EoS at high baryon densities. Similar structure for $\Lambda$ production on the EoS is found \cite{Fe11a}. The $K^{0}/K^{+}$ ratio was proposed as a sensitive probe to extract the high-density behavior of the nuclear symmetry energy (isospin asymmetric part of EoS) \cite{Fe06,Pr10}, which is poorly known up to now but has an important application in astrophysics, such as the structure of neutron star, the cooling of protoneutron stars, the nucleosynthesis during supernova explosion of massive stars etc \cite{St05}. The ratio of $\Sigma^{-}/\Sigma^{+}$ is also sensitive to the stiffness of symmetry energy in the high-density region from the UrQMD calculations \cite{Li05}. In dynamical evolutions strange particles produced in heavy-ion collisions can be easily deviated by surrounding hadrons. Consequently, the distributions in phase space and the spectra of isospin ratios are to be modified and the constraint on the symmetry energy is also influenced.

In this work, the dynamics of strange particles in heavy-ion collisions and the high-density behavior of symmetry energy from strangeness production are to be investigated with an isospin and momentum dependent transport model (Lanzhou quantum molecular dynamics (LQMD)), in which strangeness is contributed from the channels of baryon-baryon and pion-baryon collisions \cite{Fe10}. The model has been developed to treat the nuclear dynamics at near Coulomb barrier energies and also to describe the capture of two heavy colliding nuclides to form a superheavy nucleus \cite{Fe05}. Further improvements of the LQMD model have been performed in order to investigate the dynamics of pion and strangeness productions in heavy-ion collisions and to extract the information of isospin asymmetric EoS at supra-saturation densities \cite{Fe10,Fe09}. The momentum dependence of the symmetry potential was also implemented in the model, which results in an isospin splitting of proton and neutron effective mass in nuclear medium \cite{Fe11}. We have included the resonances ($\Delta$(1232), N*(1440), N*(1535)), hyperons ($\Lambda$, $\Sigma$) and mesons ($\pi$, $K$, $\eta$) in hadron-hadron collisions and the decays of resonances for treating heavy-ion collisions in the region of 1A GeV energies. The in-medium modifications of strangeness production and dynamical evolutions in nuclear medium are considered by distinguishing the isospin effects of the mean-field potentials and with correcting the threshold energies. The influence on collective flows and isospin emission will be investigated.

\section{Model description}

In the LQMD model, the time evolutions of the baryons (nucleons and resonances), hyperons and mesons in reaction system under a self-consistently generated mean-field are governed by Hamilton's equations of motion, which read as
\begin{eqnarray}
\dot{\mathbf{p}}_{i}=-\frac{\partial H}{\partial\mathbf{r}_{i}},
\quad \dot{\mathbf{r}}_{i}=\frac{\partial H}{\partial\mathbf{p}_{i}}.
\end{eqnarray}
The Hamiltonian of baryons consists of the relativistic energy, the effective interaction potential and the momentum dependent part as follows:
\begin{equation}
H_{B}=\sum_{i}\sqrt{\textbf{p}_{i}^{2}+m_{i}^{2}}+U_{int}+U_{mom}.
\end{equation}
Here the $\textbf{p}_{i}$ and $m_{i}$ represent the momentum and the mass of the baryons. The effective interaction potential $U_{int}$ is composed of the Coulomb interaction and the local potential \cite{Fe11}. The local interaction potential is expressed through the energy-density functional as the form of
$U_{loc}=\int V_{loc}(\rho(\mathbf{r}))d\mathbf{r}$. The energy-density functional reads
\begin{equation}
V_{loc}(\rho)= \frac{\alpha}{2}\frac{\rho^{2}}{\rho_{0}}+
\frac{\beta}{1+\gamma}\frac{\rho^{1+\gamma}}{\rho_{0}^{\gamma}} + E_{sym}^{loc}(\rho)\rho\delta^{2} + \frac{g_{sur}}{2\rho_{0}}(\nabla\rho)^{2} + \frac{g_{sur}^{iso}}{2\rho_{0}}[\nabla(\rho_{n}-\rho_{p})]^{2},
\end{equation}
where the $\rho_{n}$, $\rho_{p}$ and $\rho=\rho_{n}+\rho_{p}$ are the neutron, proton and total densities, respectively, and the $\delta=(\rho_{n}-\rho_{p})/(\rho_{n}+\rho_{p})$ being the isospin asymmetry. The coefficients $\alpha$, $\beta$, $\gamma$, $g_{sur}$, $g_{sur}^{iso}$ and $\rho_{0}$ are set to be the values of -215.7 MeV, 142.4 MeV, 1.322, 23 MeV fm$^{2}$, -2.7 MeV fm$^{2}$ and 0.16 fm$^{-3}$, respectively. A compression modulus of K=230 MeV for isospin symmetric nuclear matter is obtained with these parameters. Here we have moved away the bulk contribution of effective mass term in Skyrme energy-density functional, which enhances the stability in initialization and is of importance in low-energy nuclear dynamics \cite{Fe05}. But in intermediate- and high-energy heavy-ion collisions, the term has a neglectable contribution in nuclear dynamics if the same incompressible modulus is given. The nucleon effective mass in nuclear medium is contributed from the momentum dependent interaction in the LQMD model. A Skyrme-type momentum-dependent potential is used in the LQMD model \cite{Fe11}
\begin{eqnarray}
U_{mom}=&& \frac{1}{2\rho_{0}}\sum_{i,j,j\neq i}\sum_{\tau,\tau'}C_{\tau,\tau'}\delta_{\tau,\tau_{i}}\delta_{\tau',\tau_{j}}\int\int\int d \textbf{p}d\textbf{p}'d\textbf{r} f_{i}(\textbf{r},\textbf{p},t)    \nonumber \\
&& \times [\ln(\epsilon(\textbf{p}-\textbf{p}')^{2}+1)]^{2} f_{j}(\textbf{r},\textbf{p}',t).
\end{eqnarray}
Here $C_{\tau,\tau}=C_{mom}(1+x)$, $C_{\tau,\tau'}=C_{mom}(1-x)$ ($\tau\neq\tau'$) and the isospin symbols $\tau$($\tau'$) represent proton or neutron. The parameters $C_{mom}$ and $\epsilon$ was determined by fitting the real part of optical potential as a function of incident energy from the proton-nucleus elastic scattering data. In the calculation, we take the values of 1.76 MeV, 500 c$^{2}$/GeV$^{2}$ for the $C_{mom}$ and $\epsilon$, respectively, which result in the effective mass $m^{\ast}/m$=0.75 in nuclear medium at saturation density for symmetric nuclear matter. The parameter $x$ as the strength of the isospin splitting with the value of -0.65 is taken in this work, which has the mass splitting of $m^{\ast}_{n}>m^{\ast}_{p}$ in nuclear medium.

The symmetry energy is composed of three parts, namely the kinetic energy from fermionic motion, the local density-dependent interaction and the momentum-dependent potential as
\begin{equation}
E_{sym}(\rho)=\frac{1}{3}\frac{\hbar^{2}}{2m}\left(\frac{3}{2}\pi^{2}\rho\right)^{2/3}+E_{sym}^{loc}(\rho)+E_{sym}^{mom}(\rho).
\end{equation}
The local part is adjusted to mimic predictions of the symmetry energy calculated by microscopical or phenomenological many-body theories and has two-type forms as follows:
\begin{equation}
E_{sym}^{loc}(\rho)=\frac{1}{2}C_{sym}(\rho/\rho_{0})^{\gamma_{s}},
\end{equation}
and
\begin{equation}
E_{sym}^{loc}(\rho)=a_{sym}(\rho/\rho_{0})+b_{sym}(\rho/\rho_{0})^{2}.
\end{equation}
The parameters $C_{sym}$, $a_{sym}$ and $b_{sym}$ are taken as the values of 52.5 MeV, 43 MeV, -16.75 MeV. The values of $\gamma_{s}$=0.5, 1., 2. have the soft, linear and hard symmetry energy with baryon density, respectively, and the Eq. (7) gives a supersoft symmetry energy, which cover the largely uncertain of nuclear symmetry energy, particularly at supra-saturation densities. Shown in Fig. 1 is a comparison of symmetry energy for different stiffness as a function of baryon density. All cases cross at saturation density with the value of 31.5 MeV, which are basically consistent with the standard Skyrme-Hartree-Fock \cite{Br85,Ch97} and liquid-drop model or its extension calculations \cite{Mo95}. But the difference is large in the domain of high densities. In the calculation, we chose two typical cases, i.e., hard and soft symmetry energies. The high-density behavior of symmetry energy can be constrained with isospin sensitive observables calculated by transport models in comparison to experimental data.

\begin{figure}
\begin{center}
{\includegraphics[width=1.\textwidth]{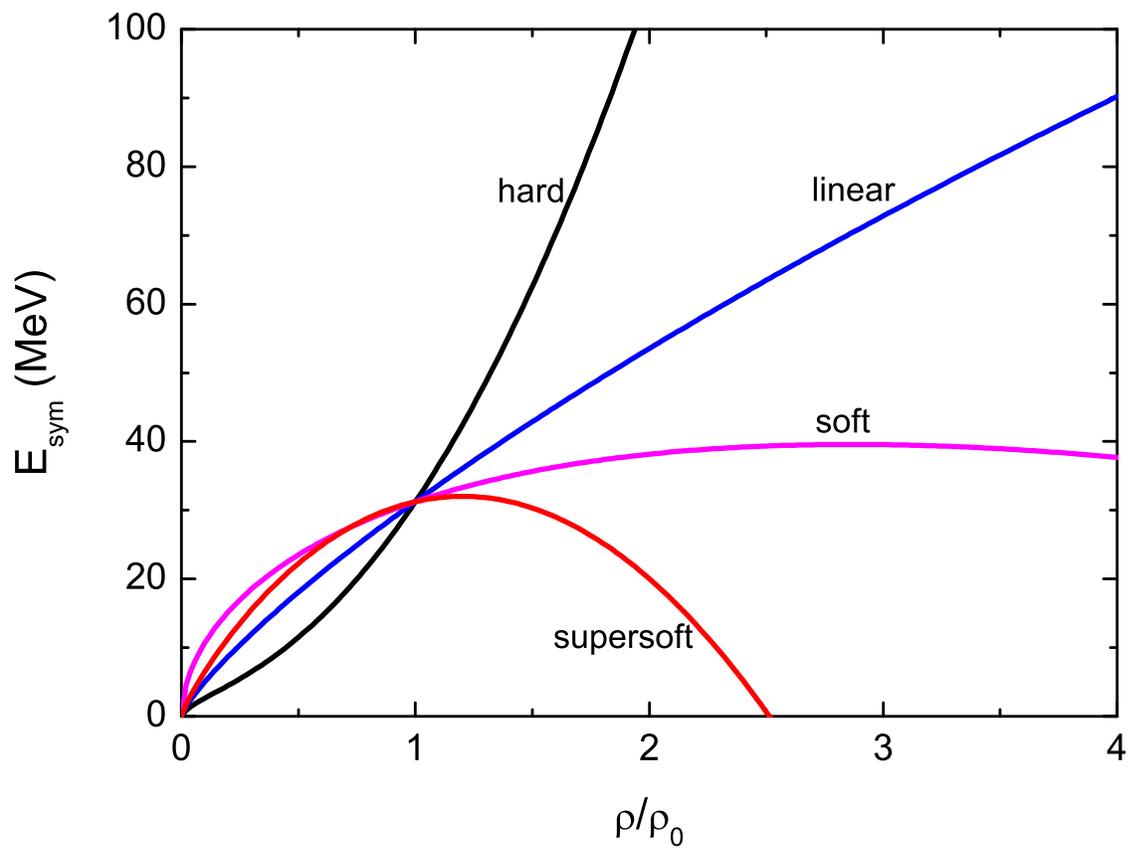}}
\end{center}
\caption{Density dependence of the nuclear symmetry energy for the cases of supersoft, soft, linear and hard at supra-saturation densities.}
\end{figure}

The hyperon mean-field potential is constructed on the basis of the light-quark counting rule. The self-energies of hyperons are assumed to be two thirds of that experienced by nucleons. Thus, the in-medium dispersion relation reads
\begin{equation}
\omega(\textbf{p}_{i},\rho_{i})=\sqrt{(m_{H}+\Sigma_{S}^{H})^{2}+\textbf{p}_{i}^{2}} + \Sigma_{V}^{H}
\end{equation}
with $\Sigma_{S}^{H}= 2 \Sigma_{S}^{N}/3$ and $\Sigma_{V}^{H}= 2 \Sigma_{V}^{N}/3$, which leads to the hyperon-nucleon (HN) potential in nuclear matter at the saturation density being the value of -32 MeV. The nuclear scalar $\Sigma_{S}^{N}$ and vector $\Sigma_{V}^{N}$ self-energies are computed from the relativistic mean-field model with the NL3 parameter ($g_{\sigma N}^{2}$=80.82, $g_{\omega N}^{2}$=155). Shown in Fig. 2 is the $\Lambda$ energy in units of free mass and optical potential in nuclear medium. The empirical value extracted from hypernuclei experiments \cite{Mi88} can be well reproduced by the optical potential. The same approach is used for the $\Sigma$ propagation in nuclear medium.

\begin{figure}
\begin{center}
{\includegraphics[width=1.\textwidth]{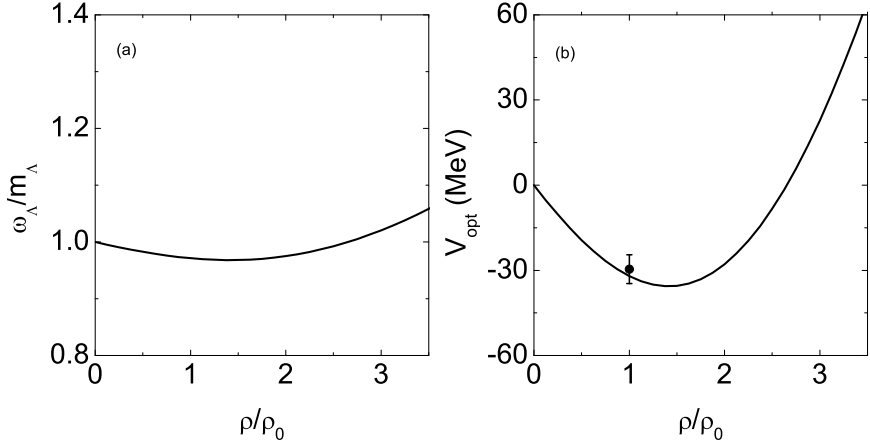}}
\end{center}
\caption{The lambda energy in units of free mass and optical potential as a function of baryon density. The empirical value extracted from hypernuclei experiments is denoted by the solid circle \cite{Mi88}.}
\end{figure}

The evolution of mesons (here mainly pions and kaons) is also determined by the Hamiltonian, which is given by
\begin{eqnarray}
H_{M}&& = \sum_{i=1}^{N_{M}}\left( V_{i}^{\textrm{Coul}} + \omega(\textbf{p}_{i},\rho_{i}) \right).
\end{eqnarray}
Here the Coulomb interaction is given by
\begin{equation}
V_{i}^{\textrm{Coul}}=\sum_{j=1}^{N_{B}}\frac{e_{i}e_{j}}{r_{ij}},
\end{equation}
where the $N_{M}$ and $N_{B}$ are the total numbers of mesons and baryons including charged resonances. Starting from the chiral Lagrangian the Klein-Gordon equations for kaon (K$^{0}$, K$^{+}$) and anti-kaon ($\overline{K}^{0}$, K$^{-}$) in nuclear medium from the Euler-Lagrange equations \cite{Li97,Ka86}
\begin{equation}
\left[\partial_{\mu}\partial^{\mu}\pm\frac{3i}{4f_{\pi}^{\ast 2}}j_{\mu}\partial^{\mu}+\left(m_{K}^{2}-\frac{\Sigma_{KN}}{f_{\pi}^{2}}
\rho_{s}\right)\right]\phi_{K}(x)=0.
\end{equation}
Here $j_{\mu}$ and $\rho_{s}$ are the baryon four-vector current and the scalar density, respectively. $K$ refers kaon and anti-kaon mesons, $\Sigma_{KN}$ the kaon-nucleon sigma term being the value in the region 200-450 MeV, $f_{\pi}$=93 MeV the pion decay constant and $f_{\pi}^{\ast}$ the in-medium pion decay constant. The above equation can be rewritten in the form as
\begin{equation}
[(\partial_{\mu}\pm iV_{\mu})^{2}+m_{K}^{\ast 2}]\phi_{K}(x)=0.
\end{equation}
In isospin asymmetric nuclear matter, the vector potential is defined by \cite{Sc97}
\begin{equation}
V_{\mu}=\frac{3}{8f_{\pi}^{\ast 2}}j_{\mu}+\tau_{3}\frac{1}{8f_{\pi}^{\ast 2}}j_{\mu 3}.
\end{equation}
The effective mass of kaon is given by
\begin{equation}
m_{K}^{\ast 2}=m_{K}^{2}-\frac{\Sigma_{KN}}{f_{\pi}^{2}}\rho_{s}-
\tau_{3}\frac{C}{f_{\pi}^{2}}\rho_{s3}+V_{\mu}V^{\mu}.
\end{equation}
Here $j_{\mu 3}$ and $\rho_{s3}$ being the isovector current and the isovector scalar density, $\tau_{3}$=1 and -1 for K$^{+}$($\overline{K}^{0}$) and K$^{0}$(K$^{-}$), respectively. The isospin parameter $C$ is taken the value of 33.5 MeV. Thus, the kaon and anti-kaon energies in the nuclear medium can be written as
\begin{equation}
\omega_{K}(\textbf{p}_{i},\rho_{i})=\left[m_{K}^{2}+\textbf{p}_{i}^{2}-a_{K}\rho_{i}^{S}
-\tau_{3}c_{K}\rho_{i3}^{S}+(b_{K}\rho_{i}+\tau_{3}d_{K}\rho_{i3})^{2}\right]^{1/2}+b_{K}\rho_{i}
+\tau_{3}d_{K}\rho_{i3}
\end{equation}
and
\begin{equation}
\omega_{\overline{K}}(\textbf{p}_{i},\rho_{i})=\left[m_{\overline{K}}^{2}+\textbf{p}_{i}^{2}-a_{\overline{K}}\rho_{i}^{S}
-\tau_{3}c_{K}\rho_{i3}^{S}+(b_{K}\rho_{i}+\tau_{3}d_{K}\rho_{i3})^{2}\right]^{1/2}-b_{K}\rho_{i}
-\tau_{3}d_{K}\rho_{i3},
\end{equation}
respectively. Here the $b_{K}=3/(8f_{\pi}^{\ast 2})\approx$0.333 GeVfm$^{3}$ with assuming $f_{\pi}^{\ast}=f_{\pi}$, the $a_{K}$ and $a_{\overline{K}}$ are 0.18 GeV$^{2}$fm$^{3}$ and 0.31 GeV$^{2}$fm$^{3}$, respectively, which result in the strengths of repulsive kaon-nucleon (KN) potential and of attractive antikaon-nucleon potential with the values of 27.8 MeV and -100.3 MeV at saturation baryon density for isospin symmetric matter, respectively. The parameters $c_{K}$=0.0298 GeV$^{2}$fm$^{3}$ and $d_{K}$=0.111 GeVfm$^{3}$ determine the isospin splitting of kaons in neutron-rich nuclear matter. The optical potential of kaon is derived from the in-medium energy as $V_{opt}(\textbf{p},\rho)=\omega(\textbf{p},\rho)-\sqrt{\textbf{p}^{2}+m_{K}^{2}}$. The values of $m^{\ast}_{K}/m_{K}$=1.056 and $m^{\ast}_{\overline{K}}/m_{\overline{K}}$=0.797 at normal baryon density are concluded with the parameters in isospin symmetric nuclear matter. The effective mass is used to evaluate the threshold energy for kaon and antikaon production, e.g., the threshold energy in the pion-baryon collisions $\sqrt{s_{th}}=m^{\ast}_{Y} + m^{\ast}_{K}$. Shown in Fig. 3 is a comparison of the kaon energy in units of its free mass and the optical potential at the momentum of $\textbf{p}$=0 as a function of baryon density. In neutron-rich nuclear matter, an isospin splitting for the pairs (K$^{0}$, K$^{+}$) and ($\overline{K}^{0}$, K$^{-}$) is pronounced with the baryon density. The equations of motion include the spatial component of the vector potential, which would lead to an attractive Lorentz force between kaons and nucleons as \cite{Zh04}
\begin{equation}
\frac{d\mathbf{p}_{i}}{dt}=-\frac{\partial V_{i}^{\textrm{Coul}}}{\partial\mathbf{r}_{i}}-
\frac{\partial\omega_{K(\overline{K})}(\textbf{p}_{i},\rho_{i})}{\partial\mathbf{r}_{i}}\pm
\textbf{v}_{i}\frac{\partial \textbf{V}_{i}}{\partial\mathbf{r}_{i}},
\end{equation}
where $\textbf{v}_{i}$ is the velocity of kaons (anti-kaons), and $\pm$ for K and $\overline{K}$, respectively. The Lorentz force increases with the kaon momentum. However this contradicts the predictions from the theoretical studies which show an opposite momentum dependence \cite{Sh92}.

\begin{figure}
\begin{center}
{\includegraphics[width=1.\textwidth]{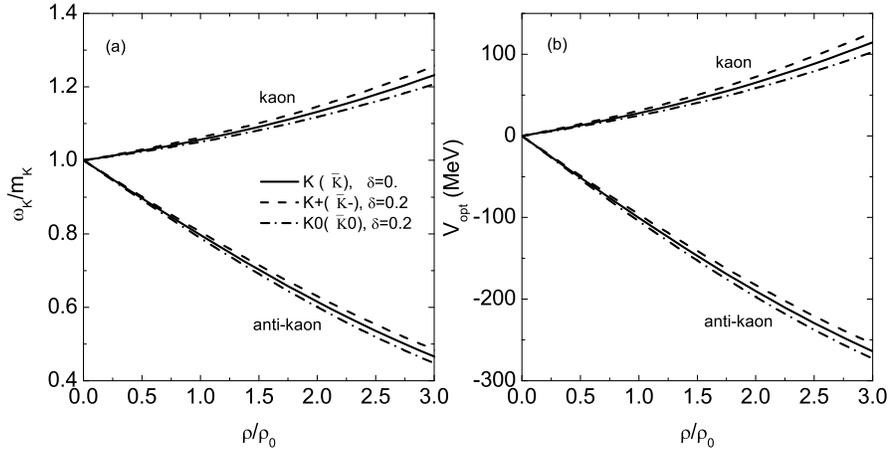}}
\end{center}
\caption{Density dependence of the kaon energy in units of free mass and the optical potential at the momentum of $\textbf{p}$=0 in isospin symmetric and asymmetric nuclear matter ($\delta$=0.2).}
\end{figure}

The scattering in two-particle collisions is performed by using a Monte Carlo procedure, in which the probability to be a channel in a collision is calculated by its contribution of the channel cross section to the total cross section. The primary products in nucleon-nucleon (NN) collisions in the region of 1\emph{A} GeV energies are the resonances of $\Delta$(1232), $N^{\ast}$(1440), $N^{\ast}$(1535) and the pions. We have included the reaction channels as follows:
\begin{eqnarray}
&& NN \leftrightarrow N\triangle, \quad  NN \leftrightarrow NN^{\ast}, \quad  NN
\leftrightarrow \triangle\triangle, \quad  \Delta \leftrightarrow N\pi, \nonumber \\
&&  N^{\ast} \leftrightarrow N\pi, \quad  NN \leftrightarrow NN\pi (s-state),  \quad N^{\ast}(1535) \rightarrow N\eta.
\end{eqnarray}
The cross sections of each channel to produce resonances are
parameterized by fitting the data calculated with the one-boson
exchange model \cite{Hu94}. At the considered energies, there are mostly $\Delta$ resonances which disintegrate into a $\pi$ and a nucleon in the evolutions. However, the $N^{\ast}$ yet gives considerable contribution to the energetic pion yields. The energy and momentum-dependent decay widths are used in the model for the resonances of $\Delta$(1232) and $N^{\ast}$(1440) \cite{Fe09}. We have taken a constant width of $\Gamma$=150 MeV for the $N^{\ast}$(1535) decay.

The strangeness is created in inelastic hadron-hadron collisions. We included the channels as follows:
\begin{eqnarray}
&& BB \rightarrow BYK,  BB \rightarrow BBK\overline{K},  B\pi \rightarrow YK,  YK \rightarrow B\pi, B\pi \rightarrow NK\overline{K}, \nonumber \\
&& Y\pi \rightarrow B\overline{K}, \quad  B\overline{K} \rightarrow Y\pi, \quad YN \rightarrow \overline{K}NN.
\end{eqnarray}
Here the B stands for (N, $\triangle$, N$^{\ast}$) and Y($\Lambda$, $\Sigma$), K(K$^{0}$, K$^{+}$) and $\overline{K}$($\overline{K}^{0}$, K$^{-}$). The parameterized cross sections of each isospin channel $BB \rightarrow BYK$ \cite{Ts99} are used in the calculation. We take the parametrizations of the channels $B\pi \rightarrow YK$ \cite{Ts94} besides the $N\pi \rightarrow \Lambda K$ reaction \cite{Cu84}. The results are close to the experimental data at near threshold energies. The cross section of antikaon production in inelastic hadron-hadron collisions is taken as the same form of the parametrization used in the hadron string dynamics (HSD) calculations \cite{Ca97}. Furthermore, the elastic scattering between strangeness and baryons are considered through the channels of $KB \rightarrow KB$, $YB \rightarrow YB$ and $\overline{K}B \rightarrow \overline{K}B$ and we use the parametrizations in Ref. \cite{Cu90}. The charge-exchange reactions between the $KN \rightarrow KN$ and $YN \rightarrow YN$ channels are included by using the same cross sections with the elastic scattering, such as $K^{0}p\rightarrow K^{+}n$, $K^{+}n\rightarrow K^{0}p$ etc.

\section{Results and discussions}

The in-medium modifications of strangeness in heavy-ion collisions can be explored from the yield distribution in momentum space. We compared the rapidity distribution of $K$ and $\overline{K}$ produced in the near-central $^{58}$Ni+$^{58}$Ni collisions at incident energy of 1.93\emph{A} GeV as shown in Fig. 4. The same as in the $^{197}$Au+$^{197}$Au reaction in Ref. \cite{Fe13}, the inclusion of the KN potential leads to a reduction of the mid-rapidity kaon yields and decreases the total multiplicity about 10$\%$. However, the antikaon-nucleon potential increases the antikaon production due to its attractive interaction in nuclear medium, which leads to a decrease of the effective mass for antikaons and dramatically enhances the antikaon yields about 200$\%$. Corrections of effective mass of kaons and antikaons in nuclear medium on the elementary cross sections are considered through the threshold energy. The effect of the Lorentz force (LF) has a slight contribution on the rapidity distribution. The experimental data are from the FOPI collaboration for K$^{+}$ denoted by solid squares \cite{Be97} and from the KaoS collaboration for K$^{+}$ and K$^{-}$ production by full circles \cite{Me00}. The open symbols are their reflections with respect to the midrapidity. The neutral strange particles K$^{0}$ and $\Lambda$ are from Ref. \cite{Me07}. In analogue to the KN interaction, the hyperon-nucleon potential reduces the hyperon yields in the mid-rapidity region as shown in Fig. 5, which is caused from the fact that the production of hyperon is associated with the kaon emission in hadron-hadron collisions, e.g., $NN \rightarrow NKY$, $\pi N \rightarrow KY$, and also the secondary collisions of hyperons with the surrounding pions in nuclear medium. Production of $\Sigma^{0}$ is suppressed about 4 with respect to the $\Lambda$ yields due to a higher threshold energy.

\begin{figure}
\begin{center}
{\includegraphics[width=1.\textwidth]{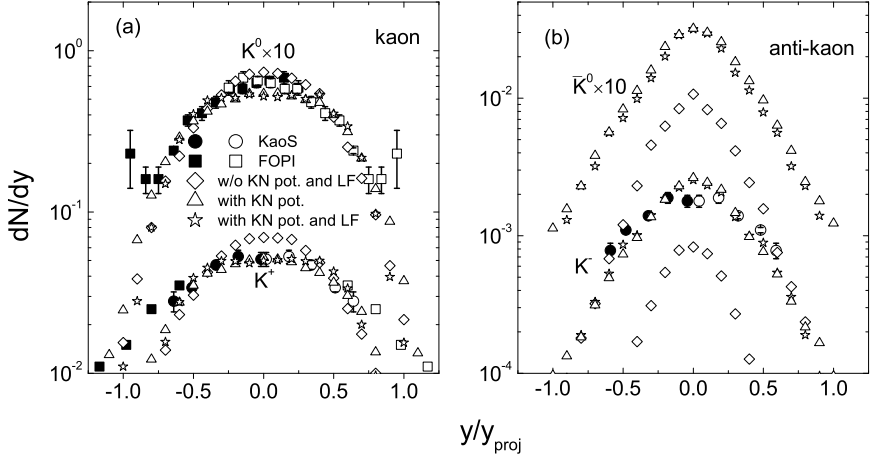}}
\end{center}
\caption{Comparison of rapidity distributions of kaons and anti-kaons in the $^{58}$Ni+$^{58}$Ni collisions at the incident energy of 1.93\emph{A} GeV with the impact parameters of b$\leq$4 for different contributions of the kaon-nucleon (KN) potential and the Lorentz-force (LF). The experimental data are from FOPI \cite{Be97} and KaoS collaborations \cite{Me00}.}
\end{figure}

\begin{figure}
\begin{center}
{\includegraphics[width=1.\textwidth]{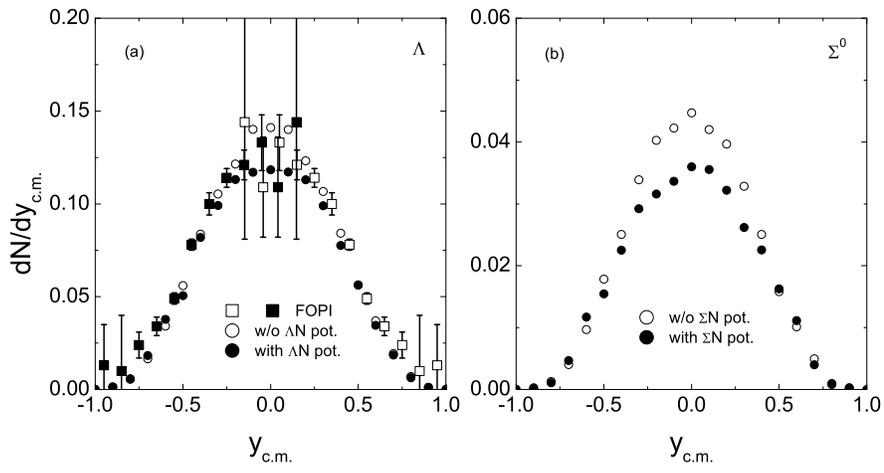}}
\end{center}
\caption{The same as in Fig. 4, but for $\Lambda$ and $\Sigma^{0}$ production.}
\end{figure}

Strangeness is produced at the compressed stage in two colliding nuclei and the secondary collisions may take place during the expansion process, in which the evolution of strangeness is deviated by surrounding baryons. The effect of the mean-field potential is pronounced from the transverse momentum distribution as shown in Fig. 6. The KN potential enhances the high-momentum kaon yields, but reduces the low-momentum part. However, the antikaon-nucleon interaction exhibits an opposite contribution. The hyperon-nucleon potential has a slight effect on the spectra, which contributes an attractive interaction at the considered energies. More pronounced observable to extract the fireball information formed in heavy-ion collisions can be obtained from the kinetic energy spectra of the invariant production cross sections as shown in Fig. 7. The invariant cross section can be fitted by the Maxwell-Boltzmann distribution as
\begin{equation}
\frac{Ed\sigma}{p^{2}dp}=C E\exp(-E_{kin}/T),
\end{equation}
where $E$ and $E_{kin}$ are the total energy and the kinetic energy, respectively, and the normalization $C$ and temperature $T$ being the fitting parameters. The inverse slope parameters $T$ of 102$\pm$10 MeV, 122$\pm$6 MeV, 124$\pm$12 MeV and 100$\pm$8 MeV are obtained for $K^{+}$, $\Lambda$, $\Sigma^{0}$ and $K^{-}$ without the optical potentials, respectively. The mean-field potential enhances the local temperatures from kaon production with 138$\pm$8 MeV and 132$\pm$6 MeV for the cases of the KN potential and the LF, respectively. However, it contributes an opposite trend for antikaon production with the values of 94$\pm$10 MeV and 82$\pm$8 MeV for the antikaon-nucleon potential and the LF, respectively. The in-medium antikaon production has a lower temperature than the kaon and hyperon emission in heavy-ion collisions, which comes from the secondary collisions, e.g., $Y\pi \leftrightarrow BK^{-}$. The conclusions are basically consistent with the experimental measurements \cite{Fo07,Me07}.

\begin{figure}
\begin{center}
{\includegraphics[width=1.5\textwidth]{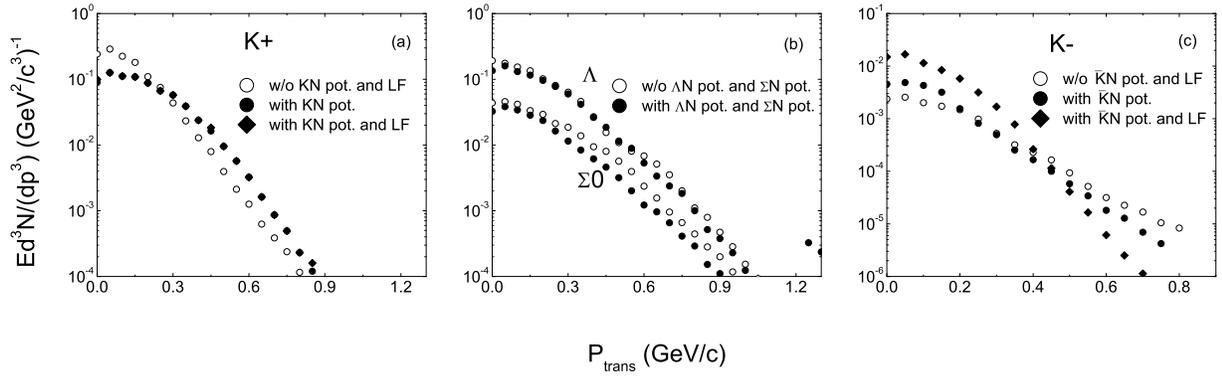}}
\end{center}
\caption{Transverse momentum distributions of strange particles produced in the 1.93\emph{A} GeV $^{58}$Ni+$^{58}$Ni collisions at the impact parameters of b$\leq$4 fm.}
\end{figure}

\begin{figure}
\begin{center}
{\includegraphics[width=1.5\textwidth]{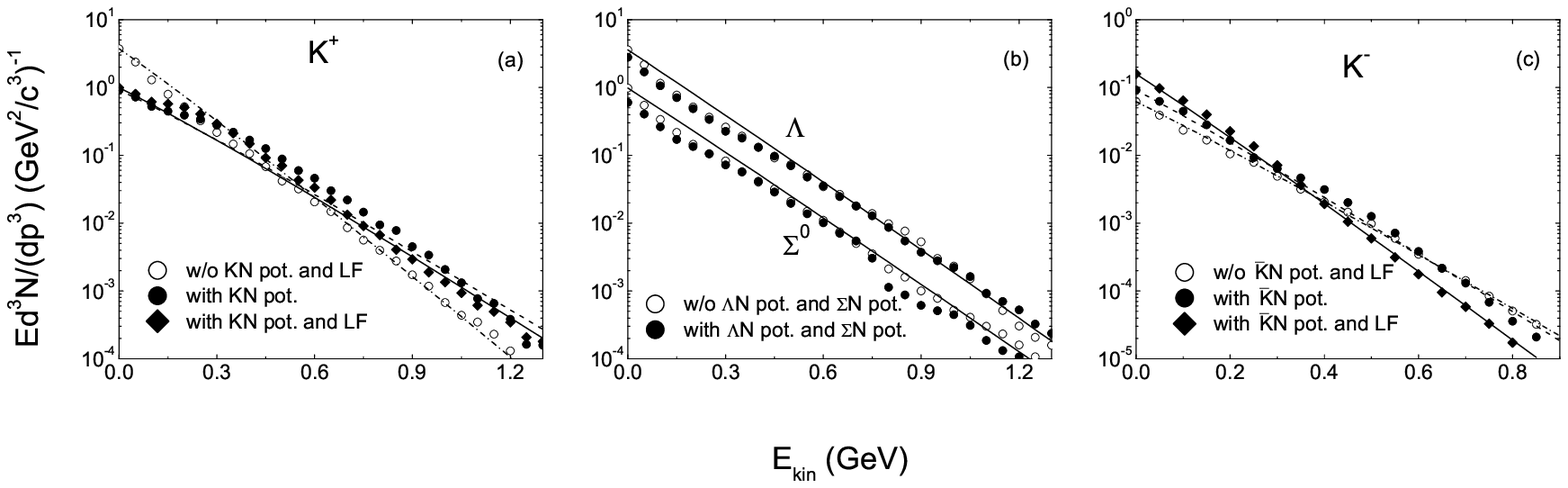}}
\end{center}
\caption{The kinetic energy spectra of inclusive invariant cross sections for the production of strange particles in collisions of $^{58}$Ni+$^{58}$Ni at the incident energy of 1.93\emph{A} GeV. The lines are fitting from the Maxwell-Boltzmann distribution.}
\end{figure}

The studies of the transverse flows of nucleons, light fragments and pions in heavy-ion collisions have motivated a lot of issues, such as the symmetry energy, in-medium NN cross section, pion-nucleon potential etc \cite{Fe12}. It has been confirmed that the in-plane flows of secondary particles are sensitive to the in-medium modifications \cite{Li95,Wa98}. To investigate strangeness production in the reaction plane and its correlation to the in-medium and relativistic effects, we computed the rapidity distributions of transverse flows for strangeness production in the near central $^{58}$Ni+$^{58}$Ni collisions (b$\leq$4 fm) at the incident energy of 1.93\emph{A} GeV. Shown in Fig. 8 is a comparison of the KN potential and LF contribution for the K$^{0}$ and K$^{+}$ production. One notices that the repulsive KN potential tends to push the kaons away from the spectator nucleons, and even leads to the appearance of antiflows. The effect of the LF in the kaon dynamics gives an attractive interaction and reduces the contribution of KN potential. Finally, a flat structure is obtained and close to the FOPI data \cite{He99}. The LF contribution to the kaon flow is also studied in Ref. \cite{Zh04} and the same conclusions are drawn. The azimuthal anisotropy of kaon emission is also studied from the directed flow in the $^{197}$Au+$^{197}$Au reaction at an incident energy of 1.5\emph{A} GeV as shown in Fig. 9. The same trends with the transverse flow in the $^{58}$Ni+$^{58}$Ni reaction are concluded. The case without the KN potential is close to the proton flows. The inclusion of LF leads to a flat distribution in the mid-rapidity region. The transverse flows of $\Lambda$ and $\Sigma^{0}$ are also calculated as shown in Fig. 10. The hyperon-nucleon potential enlarges the high-rapidity hyperon emission, which presents an opposite contribution in comparison with the KN potential. The FOPI data for the $\Lambda$ in-plane flow \cite{Ri95} are well reproduced with the inclusion of $\Lambda$N potential. The similar structure is found for the $\Sigma^{0}$ flows. The full symbols are the measured data and the open ones are reflected with respect to the midrapidity. The hyperons experience an attractive mean-field in nuclear medium and follow the nucleon flows in heavy-ion collisions.

\begin{figure}
\begin{center}
{\includegraphics[width=1.\textwidth]{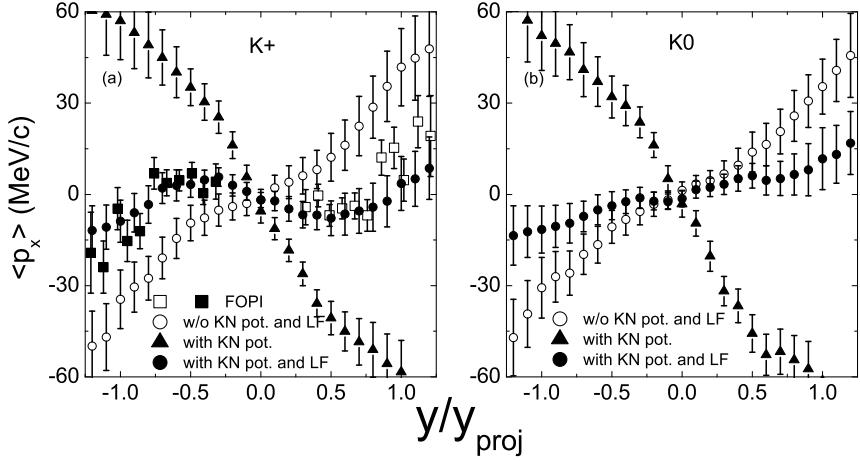}}
\end{center}
\caption{Transverse flows of kaons as a function of rapidity in the $^{58}$Ni+$^{58}$Ni reactions at the incident energy of 1.93\emph{A} GeV with different contributions of the kaon-nucleon (KN) potential and the Lorentz-force (LF). The full and open squares are the FOPI data and their reflections with respect to the midrapidity \cite{He99}.}
\end{figure}

\begin{figure}
\begin{center}
{\includegraphics[width=1.\textwidth]{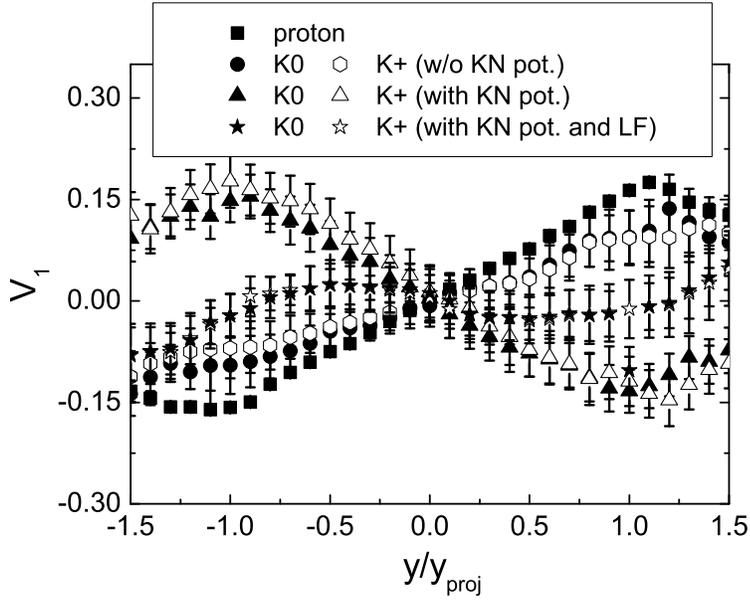}}
\end{center}
\caption{Directed flows of kaons produced in the $^{197}$Au+$^{197}$Au reaction at an incident energy of 1.5\emph{A} GeV as a function of the longitudinal rapidity. Proton flows are also implemented for a comparison.}
\end{figure}

\begin{figure}
\begin{center}
{\includegraphics[width=1.\textwidth]{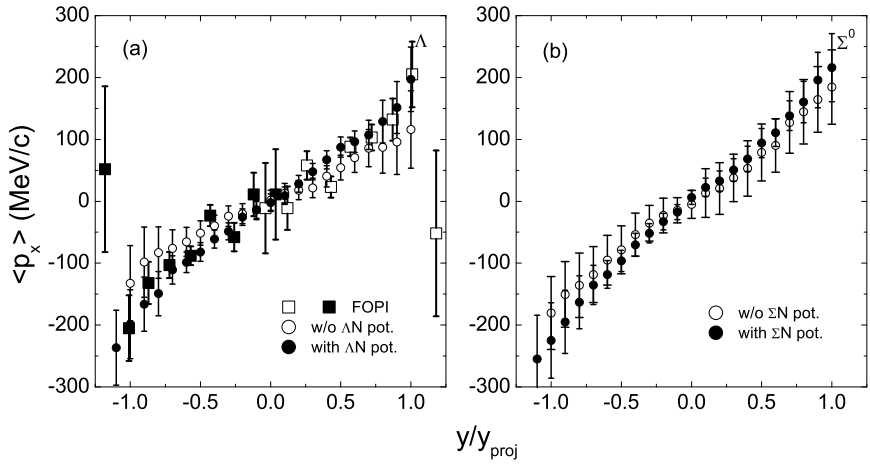}}
\end{center}
\caption{The same as in Fig. 8, but for neutral hyperon ($\Lambda$ and $\Sigma^{0}$) production. The experimental data are measured by the FOPI collaboration \cite{Ri95}}
\end{figure}

The high-density matter larger than normal nuclear density can be formed in high-energy heavy-ion collisions, where strange particles are produced. The influence of the in-medium potential by distinguishing isospin effect and the stiffness of symmetry energy on strangeness production can be observed from the emission of isospin pairs. Shown in Fig. 11 is a comparison of the energy dependence of the yields of K$^{0}$ and K$^{+}$ produced in central $^{197}$Au+$^{197}$Au collisions with and without distinguishing the isospin effects of the KN potentials in the left window. The effect of symmetry energy is shown from the ratio of K$^{0}$/K$^{+}$ in the right window. It should be noticed that the KN potential by distinguishing isospin effect increases the K$^{0}$/K$^{+}$ ratio and also enlarges the influence of symmetry energy on isospin ratio in comparison to the in-vacuum and the nonisospin cases, in particular in the domain of deep subthreshold energies. The effect of symmetry energy on the isospin ratios disappears at high incident energy, i.e., 1.5\emph{A} GeV and 1.8\emph{A} GeV. The inclusion of the KN potential leads to a reduction of the total kaon yields. The contribution of LF is very small on the kaon production, but changes the kaon distribution in momentum space. The high-density behavior of nuclear symmetry energy can be also investigated from the ratio of $\Sigma^{-}/\Sigma^{+}$ as shown in Fig. 12. In analogue to the isospin ratio of kaons, the effect of symmetry energy is pronounced in the region of the deep threshold energies. Overall, a hard symmetry energy leads to a lower isospin ratio in neutron-rich nuclear collisions. Production of sigma increases dramatically with the incident energy and the isospin effects of the yields for charged sigmas are obvious with decreasing the energy. In despite of part of sigma can be absorbed in nuclear matter through the reaction of $\pi \Sigma \rightarrow \overline{K}B$. Most of $\Sigma$ are produced in the high-density region in heavy-ion collisions, which can be used to probe the information of the high-density phase diagram of nuclear matter. Shown in Fig. 13 is a comparison of the transverse momentum distributions of K$^{0}$/K$^{+}$ and $\Sigma^{-}/\Sigma^{+}$ in the $^{197}$Au+$^{197}$Au reaction at the incident energy of 1.5\emph{A} GeV. It should be noticed that the isospin effect is pronounced in the domain of high transverse momenta. The isospin ratio decreases with the transverse momentum of strangeness. The situation is consistent with the squeeze-out nucleon emissions in heavy-ion collisions \cite{Fe12}.

\begin{figure}
\begin{center}
{\includegraphics[width=1.\textwidth]{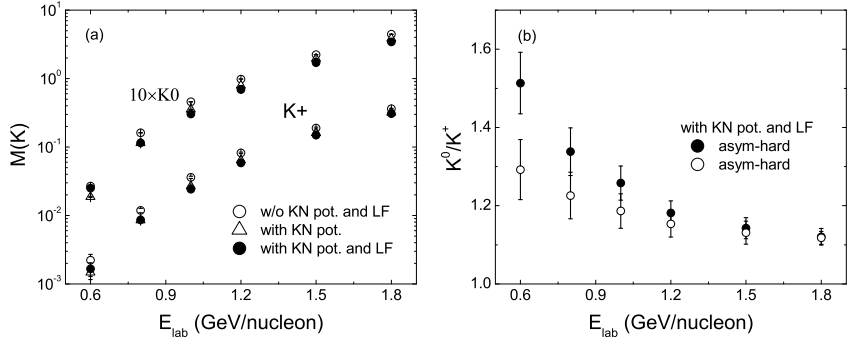}}
\end{center}
\caption{Total multiplicities of K$^{0}$ and K$^{+}$ with and without the isospin effects of the KN potentials as a function of incident energy in central $^{197}$Au+$^{197}$Au collisions for a hard symmetry energy (left panel). Excitation functions of the K$^{0}$/K$^{+}$ yields for different KN potentials (right panel).}
\end{figure}

\begin{figure}
\begin{center}
{\includegraphics[width=1.\textwidth]{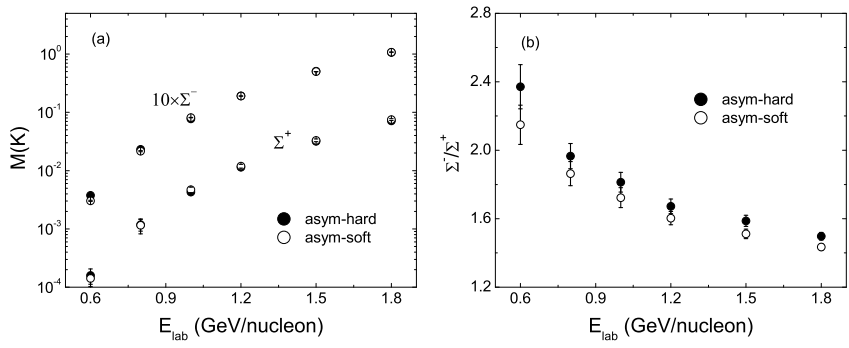}}
\end{center}
\caption{Excitation functions of the total charged sigmas and the ratio of $\Sigma^{-}/\Sigma^{+}$ in the $^{197}$Au+$^{197}$Au reaction for the hard and soft symmetry energies.}
\end{figure}

\begin{figure}
\begin{center}
{\includegraphics[width=1.\textwidth]{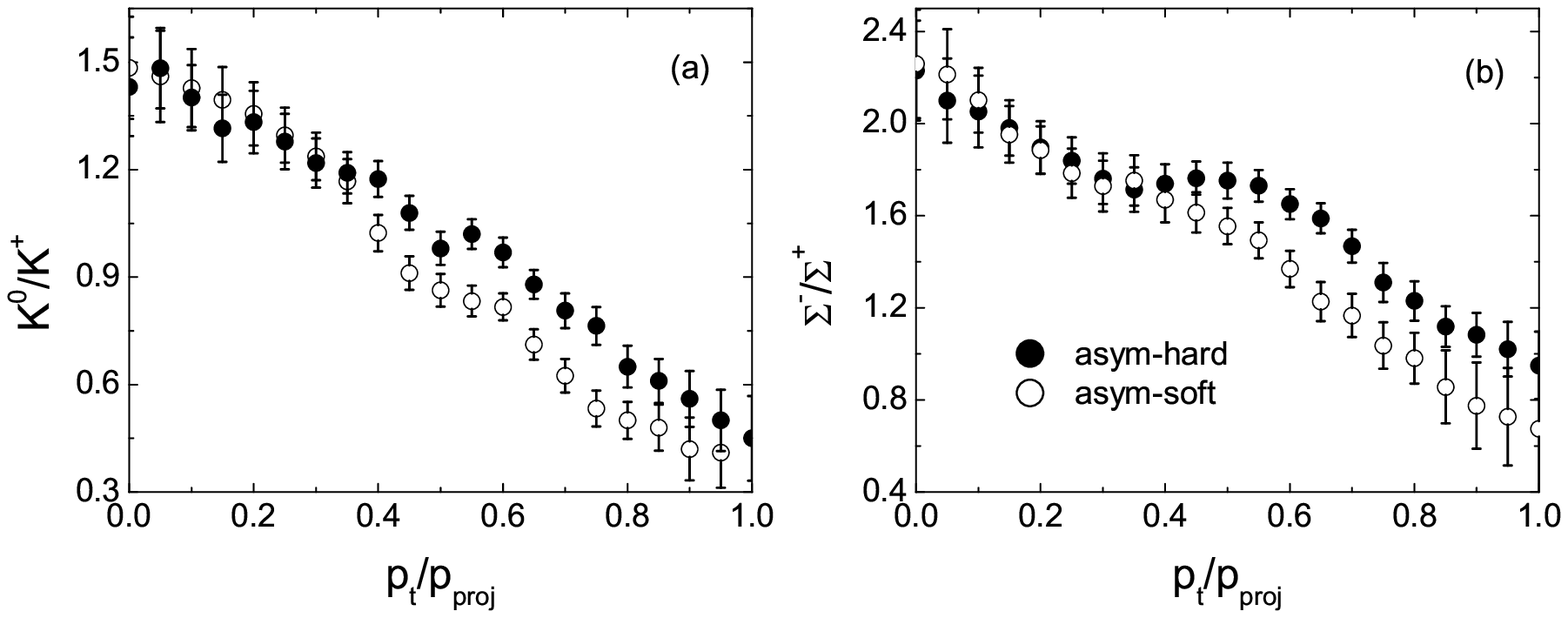}}
\end{center}
\caption{Transverse momentum distributions of the isospin ratios of K$^{0}$/K$^{+}$ and $\Sigma^{-}/\Sigma^{+}$ in the $^{197}$Au+$^{197}$Au reaction at the incident energy of 1.5\emph{A} GeV for the hard and soft symmetry energies, respectively.}
\end{figure}

\section{Conclusions}

In summary, dynamics of strange particles and collective flows in heavy-ion collisions at near threshold energies have been investigated by using the LQMD model. It is found that the in-medium potentials are of importance on strangeness emission in phase space. Specifically, the kaon-nucleon potential leads to a decrease of the kaon yields in the mid-rapidity region and also at low transverse momenta, but enhancing the production at high transverse momenta. However, the antikaon-nucleon potential exhibits an opposite contribution in nuclear medium. After inclusion of the kaon-nucleon and hyperon-nucleon potentials, the experimental data from FOPI collaboration on the transverse flows can be well reproduced with the model. The KN potential with distinguishing isospin effect enlarges the effect of symmetry energy on the ratio of $K^{0}/K^{+}$. The contribution of LF is very small on the total kaon yields, but influences the kaon emissions in heavy-ion collisions. The isospin ratios of $K^{0}/K^{+}$ and $\Sigma^{-}/\Sigma^{+}$ are sensitive to the stiffness of nuclear symmetry energy, which would be the promising probes to extract the high-density information of nuclear symmetry energy, in particular in the domain of subthreshold energies.

Strange particles in heavy-ion collisions at several GeV energies have the low yields in comparison to nucleon or pion production. It needs the high statistics to simulate strangeness production with transport models, in particular at the subthreshold energies. But the isospin effect is pronounced with decreasing the incident energy. Precise measurements on subthreshold strangeness production from the neutron-rich nuclear collisions are very necessary, which open a useful way to investigate the in-medium properties of strangeness and to constrain the high-density behavior of symmetry energy.

\textbf{Acknowledgements}

This work was supported by the National Natural Science Foundation of China under Grant Nos 11175218 and 10805061, the Youth Innovation Promotion Association of Chinese Academy of Sciences, and the Knowledge Innovation Project (KJCX2-EW-N01) of Chinese Academy of Sciences.

\end{document}